\documentclass{JHEP}

\newcommand{\be}{\begin{equation}}
\newcommand{\ee}{\end{equation}}

\newcommand{\bea}{\begin{eqnarray}}
\newcommand{\eea}{\end{eqnarray}}
\newcommand{\bean}{\begin{eqnarray*}}
\newcommand{\eean}{\end{eqnarray*}}

\def\beq{\begin{equation}}

\def\eeq{\end{equation}}

\def\Tr{\mathop{\rm Tr}}
\def\tr{\mathop{\rm tr}}

\preprint{ {\tt hep-th/0212274}}

\title{Note on Matrix Model with Massless Flavors}

\author{Bo Feng \\ Institute for Advanced Study \\
Einstein Drive, \\
Princeton, New Jersey, 08540 \\
email: fengb@sns.ias.edu
}

\abstract{In this note, following the work of Seiberg in
hep-th/0211234 for the conjecture between the  field theory
and matrix model in the case with massive fundamental flavors,
we generalize it to the  case with massless fundamental flavors.
We show that with a little 
modifications, the analysis given by Seiberg can be used directly
to the case of massless flavors.
Furthermore, this new method explains the insertion of delta functions
in the matrix model given by Demasure and Janik in hep-th/0211082.}

\begin{document}

\section{Introduction}
The field theory {\it v.s}  matrix model conjecture proposed by
Dijkgraaf and Vafa \cite{Dijkgraaf:2002fc,Dijkgraaf:2002vw,Dijkgraaf:2002dh}
has caught a lot of attentions recently(from 
\cite{Dijkgraaf:2002pp} to \cite{SOP}). The beautiful point of this
conjecture is that a lot of non-trivial complex calculations of
 exact low energy superpotential in the field
theory side can be easily done in the corresponding matrix model. Even 
in the case that there is no exact solution in the matrix model,
perturbative calculation is still much easier than the one in the
corresponding field theory. For example, the Montonen-Oliver 
duality can be derived (at least perturbatively) by matrix model
in \cite{Dijkgraaf:2002pp}\footnote{We want to thank Freddy Cachazo
for discussing about this point.}.

Although the matrix model conjecture is inspired by studies of 
geometric engineering of field theories in string theory,
it can be proved in the pure field theory. Two recent papers
 \cite{Dijkgraaf:2002xd,Cachazo:2002ry} laid solid grounds for
this conjecture. It was realized that  properties of chiral
rings and anomalies are very powerful tools to uniquely determine
the exact low energy effective superpotential. These two papers dealt with
the prototype example, i.e., ${\cal N}=1$ 
 $U(N)$ gauge theory with one adjoint
field $\Phi$ and tree level superpotential deformed from the
corresponding ${\cal N}=2$ theory. Using same idea 
in \cite{Cachazo:2002ry} Seiberg has  generalized the field theory
proof of matrix model conjecture to the 
 massive fundamental flavors \cite{Seiberg:2002jq}. 
The result explained
why field theory calculations are recaptured by 
matrix models studies in \cite{Feng:2002zb,Feng:2002yf,Demasure:2002sc,flavors}.

In \cite{Seiberg:2002jq}, it was assumed that the matrix $m(z)$ is not
degenerated, so all quarks are massive and are integrated out. 
Therefore, the low energy effective action does not depend on
the meson operator. The result is related to the fact that polynomials 
$q(z)$ has been fixed when solving the loop equation. It was also
remarked in  \cite{Seiberg:2002jq} that an ambiguity in $q(z)$ will
sign the appearance of more fields in the effective superpotential.

In this  note, we will try to do a little more than that  in  
\cite{Seiberg:2002jq} by letting matrix $m(z)$ to be degenerated.
Under the new situation, the low energy effective action will be a function of
not only glueball fields $S_i$, but also some massless fields
(mesons $M_i^j$). In fact, we will show that with just a little
bit of modifications, the method of  \cite{Seiberg:2002jq}
can be applied to our new cases.

The organization of the note is following. In section two we
briefly review the result in  \cite{Seiberg:2002jq}. In section three
we make our proposal to the study of massless flavors. In section
four we give  conclusion and discussion. 

\section{Review of results in \cite{Seiberg:2002jq}}
In \cite{Seiberg:2002jq}, Seiberg considered a general ${\cal N}=1$
$U(N)$ theory with the adjoint field $\Phi$, $N_f$ fundamental 
$Q^i$ and $N_f$ anti-fundamental $\widetilde{Q}_j$. The tree level
superpotential is  
\be
\label{Sei-1.1}
W_{tree}= \Tr W(\Phi)+ \widetilde{Q}_i m^{i}_j(\Phi) Q^j
\ee
with the function $W(\Phi)$ and matrix $m^{i}_j(\Phi)$ to be 
following polynomials
\bea
W(z) & = & \sum_{k=1}^n {g_k \over k+1} z^{k+1} \label{Phi-1.2}\\
 m^{i}_j(z) & = & \sum_{k=1}^{l+1} m_{j,k}^i z^{k-1} \label{m-matrix}
\eea
To solve the theory, it is useful to define following chiral operators
\bea
T(z) & = & \Tr \Bigl({1 \over z-\Phi}\Bigr),  \label{2.1a} \\
w_{\alpha}(z) & = & {1\over 4\pi}  \Tr  \Bigl({W_{\alpha} \over z-\Phi}\Bigr),
\label{2.1b} \\
R(z) & = & {-1\over 32\pi^2}  \Tr\Bigl({W_{\alpha} W^{\alpha} \over z-\Phi}\Bigr),
\label{2.1c} \\
M_i^j & = & \widetilde{Q}_i {1\over z-\Phi} Q^j \label{2.1d}
\eea
which will satisfy following loop equations given by anomaly considerations
\begin{eqnarray}
[W'(z) T(z)]_{-}+ \tr[m'(z) M(z)]_{-} & = & 2 R(z) T(z) +w_{\alpha}(z)
w^{\alpha}(z)  \label{2.4a} \\
~[W'(z) w_{\alpha}(z)]_{-} & = &  2 R(z) w_{\alpha}(z) \label{2.4b} \\
~ [W'(z) R(z)]_{-} & = & R(z)^2  \label{2.4c} \\
~ [(M(z) m(z))_i^j]_{-} & = & R(z) \delta_i^j  \label{2.4d} \\
~[(m(z) M(z))_i^j]_{-} & = & R(z) \delta_i^j  \label{2.4e} 
\eea
where $\tr$ is over the flavor index while $\Tr$, over the  color index.
Among these equations (\ref{2.4c}) is same as in the case of pure 
adjoint field and can be solved by
\be
\label{2.5} W'(z) R(z) +{1\over 4} f(z)= R(z)^2
\ee 
with $f(z)=-4[W'(z) R(z)]_{+}$ of degree $(n-1)$. Equations 
(\ref{2.4d}) and (\ref{2.4e}) can be written as 
\bea
(M(z) m(z)-q(z))_i^j & = &  R(z) \delta_i^j,~~~q(z)=[M(z)m(z)]_{+} 
\label{2.7a} \\
(m(z) M(z)-\widetilde{q}(z))_i^j & = &  R(z) \delta_i^j,~~~
\widetilde{q}(z)=[m(z)M(z)]_{+} 
\label{2.7b} 
\eea
with consistent condition $\widetilde{q}=m q m^{-1}$. The solution of
(\ref{2.7a}) (\ref{2.7b}) can be written as
\be
\label{2.9} M(z)=R(z) m^{-1}(z)+ q(z) m^{-1}(z)
\ee
To avoid  singularities in the solution (\ref{2.9}) when $m(z)$ has a zero
eigenvalue, $q(z)$ has been uniquely fixed to remove these singularities.

If we write $W'(z)=g_n\prod_{i=1}^n (z-a_i)$, in the case that
$m(a_i)\neq 0$ for all $a_i$, all quarks are massive and should
be integrated out. The low energy field theory is just the pure
SUSY Yang-Mills theory and the remainded massless fields are these
$S_i,\omega_\alpha^i$. Using the similar counting as in  
\cite{Dijkgraaf:2002xd,Cachazo:2002ry}, it can be shown that only genus two
and genus one can contribute. The answer for the effective 
superpotential is given by 
\be
W_{eff}(S_i,w_{\alpha}^i,N_i) =\int d^2 \psi {\cal F}_0({\it S}_i(\psi))+ 
 {\cal F}_1(S_i) \label{2.30}
\ee
with
\bea
 {\cal F}_0({\it S}_i(\psi)) & = &{\cal F}_0^{pert}+{1\over 2} 
\sum_{i} {\it S}_i^2 (\log{\Lambda^3 \over {\it S}_i}+{3\over 2})
\label{2.31a} \\
 {\cal F}_1(S_i) & = & {\cal F}_1^{pert}(S_i)\label{2.31b} 
\eea
to be  contributions of sphere and disk respectively.

The corresponding matrix model for above field theory (\ref{Sei-1.1})
is given by free energy $\hat{\cal F}$ through
\be
\label{3.2} \exp(-{\hat{N}^2 \over \hat{g}^2} \hat{\cal F})
= \int d\hat{\Phi} d \hat{Q} d\hat{\widetilde{Q}} \exp(-A)
\ee
with the action 
\be
\label{3.1} A={\hat{N} \over \hat{g}}[ \Tr W(\hat{\Phi})+
\hat{\widetilde{Q}}_i m^i_j (\hat{\Phi}) \hat{Q}^j]~.
\ee
where $\hat{\Phi}$ is  $\hat{N}\times \hat{N}$ matrix, 
$\hat{\widetilde{Q}}$, $N_f\times \hat{N}$ matrix and $\hat{Q}$,
$\hat{N}\times N_f$ matrix.
The relevent resolvents are given by
\bea
\hat{R}(z) & = & {\hat{g}\over \hat{N} } \langle 
\Tr({1\over z-\hat{\Phi}}) \rangle \label{3.3a} \\
\hat{M}_i^j(z) & = & \langle \hat{\widetilde{Q}}_i {1\over z-\hat{\Phi}}
 \hat{Q}^j \rangle \label{3.3b}
\eea
In the large $\hat{N}$ limit with fixed $N_f$, loop equations in the
matrix model become
\bea
 [W'(z) \hat{R}(z)]_{-} & = & \hat{R}(z)^2  \label{3.6a} \\
~ [(\hat{M}(z) m(z))_i^j]_{-} & = & \hat{R}(z) \delta_i^j  \label{3.6b} \\
~[(m(z) \hat{M}(z))_i^j]_{-} & = & \hat{R}(z) \delta_i^j  \label{3.6c} 
\eea
which  exactly correspond  to these in the field theory 
(\ref{2.4c}) (\ref{2.4d}) (\ref{2.4e}). From these we can identify
\be  \label{3.7}
\hat{R}(z)= \langle R(z) \rangle,~~~~~\hat{M}(z) = \langle M(z) \rangle
\ee
We need to seperate out the sphere and disk contributions to free
energy in the matrix model by $\hat{\cal F}_0=\lim_{\hat{N}\rightarrow
\infty} \hat{\cal F}$, $\hat{\cal F}_1=\lim_{\hat{N}\rightarrow
\infty} {\hat{N}\over \hat{g}} (\hat{\cal F}-\hat{\cal F}_0)$. Then
it can be shown that 
\be \label{3.10}
{\cal F}_0=\hat{\cal F}_0,~~~~~~{\cal F}_1=\hat{\cal F}_1~.
\ee
Thus it established the equivalence between the gauge theory and 
the matrix model.

\section{The degenerated matrix $m_i^j(z)$}
As mentioned in the introduction, the derivation in the section two 
is under the assumption that $m(\Phi)$ is a generic non-degenerated
matrix. If the matrix  $m(\Phi)$ is degenerated, some flavors will be
massless and the low energy effective action will involve more
variables than these $S_i,w_{\alpha}^i$. However, since the $SU(N_i)$ parts
are confined and there are only $U(1)$ parts left, the proper low
energy variables are not $Q^i,\widetilde{Q}_j$, but meson fields
$M^i_j=\widetilde{Q}_j Q^i$ which are singlets of $SU(N_i)$.

To get the effective superpotential as a function of $S_i,w_{\alpha}^i$
and $M^i_j$, we will adopt a trick used in \cite{Balasubramanian:2002tm}.
In that paper, to study the field theory of double trace deformation
\be \label{double}
W(\Phi) = {g_2\over 2} \Tr(\Phi^2) + g_4 \Tr(\Phi^4)
 + \widetilde{g}_2(\Tr(\Phi^2))^2
\ee
and the corresponding matrix model, we linearlized it by involving an
additional gauge singlet field $A$ with superpotential
\be \label{double_A}
W_{tree}= {1\over 2}( g_2+ 4\widetilde{g}_2 A) \Tr (\Phi^2)+g_4 \Tr(\Phi^4)
-\widetilde{g}_2 A^2\equiv W_{single}-\widetilde{g}_2 A^2
\ee
Now, except the last term $ -\widetilde{g}_2 A^2$ which does not include
the field $\Phi$, $W_{single}$ is in  the typical form of cases
discussed in  \cite{Dijkgraaf:2002xd,Cachazo:2002ry}. According to the
prescription in these two papers, what we need to do is just to integrate
field $\Phi$ while {\sl leaving field $A$ untouched}\footnote{We like to
thank Freddy Cachazo for emphsizing this point to us.}. In another word,
the exact effective action is given by minimizing
\be
W_{single}^{(exact)}(A,S_i)-\widetilde{g}_2 A^2
\ee 
regarding to $A$. The $W_{single}^{(exact)}$ can be calculated by field
theory or matrix model methods. If we use the matrix model method, we 
need only to use $W_{single}$ in (\ref{double_A}) to calculate the 
free energy ${\cal F}_0$ and get
\be
W_{single}^{(exact)}(A,S_i)= N_i {\partial {\cal F}_0(S_i,A) \over
\partial S_i}
\ee
where measure term has been included into ${\cal F}_0(S_i,A)$ implicitely.

Same logic tells us how to deal with the degenerated matrix $m$. What we 
need to do is to involve some new fields $X_i^j$ 
such that $\widetilde{m}(\Phi,X)$ is not degenerated. Therefore
 the analysis in the section two can be applied and corresponding matrix model
can be derived. More explicitely, the tree level superpotential will
be the form
\be
\widetilde{W}= W(\Phi,Q,\widetilde{Q},X)+ L(X)
\ee
where matrix $\widetilde{m}(\Phi,X)$ in $ W(\Phi,Q,\widetilde{Q},X)$
will be non-degenerated. Now to the part $ W(\Phi,Q,\widetilde{Q},X)$,
$X$ can be treated as  parameters and situation is reduced to the
standard one in (\ref{Sei-1.1}). It is not hard to see that every
step in \cite{Seiberg:2002jq} is valid and need no modifications:
we just integrate all $\Phi$, $Q$ and $\widetilde{Q}$ (but do
not touch $X$). 
Especially the action for the corresponding matrix model is just given 
by $ W(\Phi,Q,\widetilde{Q},X)$. From this matrix model we can 
calculate $\hat{\cal F}_{k}(S_i,w_{\alpha}^i,X),~k=0,1$. The only 
modification of above procedures is that we need to minimize the
action
\be
W^{(exact)}(\Phi,Q,\widetilde{Q},X)+ L(X)
\ee 
regarding to $X$. It is exactly this minimization  giving the final 
effective action as a function of $S_i,w_{\alpha}^i$ as well as the
meson fields.

To demonstrate our idea, let us use one simple example with only 
$N_f$ flavors and degenerated  mass matrix 
$m=\left[ \begin{array}{cc} m_{K\times K} & 0 \\ 0 & 0 \end{array} \right]$.
This simple model has been discussed in  \cite{Feng:2002zb,Feng:2002yf},
but now we use our new understanding to redo it. Involving the new
field $X_i^j$, the superpotential will become
\be \label{sup-X}
W= \sum_{i,j=1}^{K} \widetilde{Q}_i m^i_j Q^j + \sum_{a,b=1}^{N_f-K}
X_a^b( -M_b^a+ \widetilde{Q}_b Q^a) \equiv \sum_{i,j=1}^{N_f}
\widetilde{Q}_i \widetilde{m}^i_j Q^j- \tr(X M)
\ee
where $X$ can be treated as Lagrangian mulitpler, $M$ mesone fields  and 
$\widetilde{m}=\left[ \begin{array}{cc} m_{K\times K} & 0 \\ 0 & X_a^b
 \end{array} \right]$. With the new non-degenerated mass matrix 
$\widetilde{m}$, we can apply the analysis in \cite{Seiberg:2002jq}
and reduce the field theory problem to the matrix model. Using 
(\ref{3.2}) and (\ref{3.1}) the matrix integration is 
\be 
 \exp(-{\hat{N}^2 \over \hat{g}^2} \hat{\cal F})
= \int  d \hat{Q} d\hat{\widetilde{Q}}\exp(-{\hat{N} \over \hat{g}}
\hat{\widetilde{Q}}_i \widetilde{m}^i_j  \hat{Q}^j)
\ee
This matrix 
integration has been done in \cite{Feng:2002zb,Feng:2002yf} and we got
\be \label{exact-X}
W_{exact}= N_c (\det(\widetilde{m}) \Lambda^{3N_c-N_f})^{1\over N_c}
-\tr(XM)=N_c (\det(m)\det(X) \Lambda^{3N_c-N_f})^{1\over N_c}
-\tr(XM)
\ee
where the first term is got from the matrix model. Next thing we need to
do is to minimize superpotential (\ref{exact-X}) regarding to $X$. 
Differentiating ${\partial \over \partial X}$ we get
\be \label{M_X}
M= X^{-1} (\det(m)\det(X) \Lambda^{3N_c-N_f})^{1\over N_c}
\ee
Taking determinant of (\ref{M_X}) at two sides, we get
\be \label{M_X_det}
\det(X)={ (\det(M))^{ N_c\over N_f-K-N_c} \over  
(\det(m) \Lambda^{3N_c-N_f})^{N_f-K \over N_f-K-N_c}}
\ee
Putting (\ref{M_X_det}) back into (\ref{M_X}) we solve
\be \label{solve_X}
X= M^{-1} ({ \det(M) \over  \det(m) \Lambda^{3N_c-N_f}})^{1\over  N_f-K-N_c}
\ee
So finally we have
\be \label{W_exact}
W_{exact}= (N_c-(N_f-K))  ({  \det(m) \Lambda^{3N_c-N_f}\over 
 \det(M)})^{1\over  N_c-(N_f-K)}
\ee
This is exactly the  result got in \cite{Feng:2002yf} 
(notice that $K$ in this paper
should be $N_f-K$ in equation (2.5) of \cite{Feng:2002yf}). From above
calculations, we see how the dependence of $W$ on  meson fields can 
be recovered from the minimization of field $X$.

We can repeat same calculation before the glueball field $S$ has been 
integrated out. The effective action is given in  \cite{Feng:2002zb}
as
\be \label{W_S_X}
W= -N_c[ S\log( { S\over (\det(m)\det(X) \Lambda^{3N_c-N_f})^{1\over N_c}})
-S] -\tr(XM)
\ee
Minimizing it regarding to $X$, we get
\be \label{M_X_S} 
X= S M^{-1}
\ee
Putting it back into (\ref{W_S_X}) we find
\be \label{W_exact_S}
W=-(N_c-(N_f-K))[ S\log ({ S \over ( { \det(m) \Lambda^{3N_c-N_f}\over \det(M)})
^{1\over N_c-(N_f-K)}}) -S]
\ee
which is same form found in \cite{Feng:2002yf} (equation (3.5)).

There are several points we like to remark. First,  results in 
 \cite{Feng:2002zb,Feng:2002yf} are got by the insertion of proper
delta-functions while here, by involving new fields $X$. In fact,
this new method shed light on the origin of delta functions. In our new
method, we minimize $X$ after the integration in matrix model.
However, in our example the contribution of $X$ to the exact effective
action comes from ${\cal F}_1$ term only, i.e., 
$W= {\cal F}_1+L(x)+....$, where $...$ are terms do not have $X$.
Minimizing $W$ relative to $X$ is equal to minimizing $  {\cal F}_1+L(x)$
in matrix model, which in turn is equal to integrating  $X$ fields 
in the matrix model using the full action (including the $tr(XM)$ terms) 
directly.
Applying the $\int dX$ to action  (\ref{sup-X}) we get immediately
the delta function
\be \label{delta}
\delta(M_b^a- \widetilde{Q}_b Q^a)~.
\ee 
So the analysis given by Seiberg explains the proposal
made in \cite{Feng:2002yf}. The delta function (\ref{delta}) can be 
understood from another point of view: it changes the coordinate
$Q,\widetilde{Q}$ in UV theory to proper coordinate $M$ in IR\footnote{
This point of view is also given by K. Ohta in \cite{Ohta:2002rd}
where the appearance of delta function is explained from this point of
view.}. 
We want to emphasize that in general
situations (for example the baryonic deformation in superpotential), 
minimizing $X$ at the level of effective action is not
same as integrating  $X$ at the level of free energy in matrix 
model (for example, the double trace deformation in 
\cite{Balasubramanian:2002tm}), so the insertion of delta functions
 will be modified.

Second, our example (\ref{sup-X}) does not apply to the case
$N_c< N_f-K$ because in this case, the number of 
parameters $M_a^b$ is  more than the one of
 physical parameters in the moduli space\footnote{
For case $N_c< N_f-K$, parameters $M$ are
$(N_f-K)^2$ while by Higgsing mechanism, there are only
$2(N_f-K)N_c-N_c^2$ massless fields. For case $N_c\geq N_f-K$,
by Higgsing mechanism, there are 
$2(N_f-K)N_c-(N_c^2-(N_c-N_f+K)^2)=(N_f-K)^2$, so it is equal to 
the number of meson fields $M$.}. To deal with these extra 
parameters, we can involve
extra fields $Y$ as  multipliers of constraints and minimize
the effective action relative to $Y$. Another way is to treat
$X_a^b$ not totally independent to each other when minimizing the
action. No matter which method to be used, it is more involving, but
also promising to solve the matrix model with baryonic deformation.

Third, when $N_c=N_f-K$,  (\ref{W_S_X}) and (\ref{W_exact_S}) reduce to
\be
W= S\log ( { det(m) \Lambda^{3N_c-N_f}\over det(M)})
\ee
which, after minimizing $S$, gives $W=0$ and
\be
 det(m) \Lambda^{3N_c-N_f}= det(M)
\ee
as the quantum corrected moduli space \cite{Demasure:2002sc,Bena:2002tn}.

\section{Conclusions}
In this paper we made a little step along the analysis given by 
Seiberg in \cite{Seiberg:2002jq} for the field theory {\it v.s.}
matrix model conjecture in the case with fundamental flavors. 
We allow the matrix
$m(\Phi)$ to be degenerated so that the low energy effective superpotential
will depend on both the glueball fields $S_i, w_{\alpha}^i$ and other
massless fields, for example, mesons $M$. We showed that, by involving
new fields $X$ as  Langrangian multipliers, the analysis in
\cite{Seiberg:2002jq} can be translated into this new case with just
a little modification. Furthermore this new method explains the 
insertion of delta function in the matrix model integrations proposed
in  \cite{Demasure:2002sc,Feng:2002yf}.

It is obvious that there are still a lot of works needed to be done.
For example, it will be nice to generalize the discussion to
the superpotential with baryonic like deformations. The case of
multi gauge groups is also interesting because it will shed light
on various things, like chiral fields and duality. 
One of  most important things is still the investigation of dualities
in the matrix model. It is very interesting to see if the matrix model
can fascinate the study of dualities in field theories.

\section*{Acknowledgements}
We want to thank the discussion with  Vijay Balasubramanian, Freddy Cachazo,
Yang-Hui He, Min-xin Huang, Vishnu Jejjala, Asad Naqvi and Nathan Seiberg.
We want also to express our appreciation of the generous hospitality of the
High Energy Group at the University of Pennsylvania. 
This research is supported under the NSF grant PHY-0070928.

\bibliographystyle{JHEP}

\end{document}